\documentclass[preprint2]{aastex}

\usepackage{graphicx}

\newcommand{\muh}{\langle \mu \rangle}

\shorttitle{Circumplanetary Disk Truncation}
\shortauthors{Mitchell and Stewart}

\begin{document}


\title{Photoevaporation as a truncation mechanism \\ for circumplanetary disks}


\author{Tyler R. Mitchell and Glen R. Stewart}
\affil{Laboratory for Atmospheric and Space Physics \\
1234 Innovation Drive \\
 Boulder, CO 80303-7814}


\begin{abstract}
We investigate the conditions under which the regular satellites of Jupiter and Saturn formed. The final stage of giant planet accretion is thought to occur slowly over a relatively long, $10$ Myr, timescale. Gas accretion during this stage, through a completely or partially opened gap in the solar nebula, occurs slowly allowing for the condensation of ices, and incomplete differentiation, seen in the regular satellites of the giant planets. Furthermore, the dichotomy seen in the Jovian and Saturnian systems may be explained as this infall wanes or is completely shutoff as a result of gap opening or global depletion of gas in the solar nebula. We present one-dimensional simulations of circumplanetary disks that couple the viscous transport of material with the loss of mass at the disk outer edge by ultraviolet photoevaporation as well as the infall of material from the solar nebula. We find that the circumplanetary disks of these protoplanets are truncated, as a result of photoevaporation, at a range of values with the mean corresponding to $\approx 0.16$ Hill radii. These truncation radii are broadly consistent with the current locations of the regular satellite systems of Jupiter and Saturn. We also find that photoevaporation can successfully act as a clearing mechanism for circumplanetary nebulae on the potentially short timescales, $10^2-10^4$ yr, over which mass accretion from the solar nebula wanes as a result of gap opening. Such a rapid clearing of the circum-Jovian disk may be required to explain the survival of the Galilean satellites.
\end{abstract}





\section{Introduction}
\label{sec:intro}

In recent years, the impact of photoevaporation on the evolution of protoplanetary disks has received much attention \citep{johnstone98,bally98,adams04,gorti09}. Recent observations confirm that there is sufficient FUV and X-ray flux from young, solar-type stars to drive photoevaporative mass loss in the surrounding nebulae \citep{ingleby11}. Although photoevaporation has been investigated in detail in the context of planet formation, it has yet to be applied to circumplanetary disks and the formation of regular satellites. 

Recent models have looked at the combined effects of viscous evolution and photoevaporation on the evolution of protoplanetary nebulae \citep{clarke07,mitchell10}. It is tempting to apply a direct analogy between protoplanetary and protosatellite nebulae, but it would be incorrect to view a protosatellite nebula as simply a scaled down version of a protoplanetary nebula. The slow infall of material from the solar nebula onto circumplanetary disks, as proposed by \citet{canup02,canup09,ward10}, causes them to behave very differently from protoplanetary analogues. In particular, the combination of mass infall and photoevaporative mass loss can allow for quasi steady-state solutions.

The giant planets have similar ratios of the total mass contained within the satellite systems to the mass of the host planets (${\rm M_{satellites} \sim 10^{-4}\ M_{planet}}$). In general, the mass of a satellite increases with increasing distance from the planet, reaches a maximum, and then decreases with distance. A striking difference is seen in the mass distributions of the satellite systems of Jupiter and Saturn. Jupiter has four large moons of relatively the same size whereas Saturn has one large moon that contains almost all of the mass in the entire satellite system. The similarities in these systems indicate that they formed from similar processes, yet these processes have acted in such a way to produce very different outcomes. 	

The low inclination and prograde orbits of the regular satellites of giant planets indicate that they formed in situ from a circumplanetary disk. The similarity of the satellite systems of giant planets, with Jupiter in particular, to the solar system has contributed to the development of formation theories that employ a ``minimum mass subnebula'' \citep{lunine82}, analogous to the ``minimum mass solar nebula'' \citep{weidenschilling77}. Summing the mass of heavy elements contained within the satellites and augmenting it to solar composition determines the mass of this subnebula. The ``MMSN'' approach leads to a minimum mass of $\sim 0.02\ M_{\rm J}$. This however leads to a variety of problems when applied to the Jovian and Saturnian systems. 

For one, the resulting nebula is too warm to condense ices, yet, with the exception of Io, condensed volatiles are known to be a major component of the Galilean satellites. In such a dense nebula, the effect of type I migration would be strong enough to migrate Callisto into Jupiter before it had sufficient time to grow to its current size \citep{ward98}. Even if type I migration did not operate, Callisto would accrete on such a short timescale it would melt and become differentiated, which is not supported by Galileo observations \citep{anderson97}. Recent gravity measurements by the Cassini spacecraft indicate that Titan is also partially differentiated which would imply a long accretion timescale for the regular satellites of Saturn as well \citep{iess10,barr10}.

Beginning with the assumption that not all of the $0.02\ M_{\rm J}$, estimated in the ``MMSN'' approach, be present at once in the protosatellite disk, \citet{canup02} have developed a model referred to as the ``gas-starved'' disk model. Their gas-starved disk has a slow infall of material, from which the satellites are slowly built. This slow infall results in a steady-state disk with a much lower instantaneous disk mass than in the aforementioned ``MMSN'' approach. In the ``gas-starved'' disk model satellites preferentially grow in the inflow region of the disk where gas and solids are delivered from the solar nebula. Once the satellites become massive enough, they migrate toward the planet, sweeping up solids along the way. The growth and loss by migration of successive generations of satellites allows for a quasi steady-state in which the total mass contained in the satellite system is maintained at $\sim 10^{-4}\ M_{\rm p}$ \citep{canup06,sasaki10}.  When the disk mass is depleted as a result of the global depletion of the solar nebula, the radial migration of satellites is halted, only the final generation of satellites is retained.

Furthermore, the observed ice-to-rock fraction of the Jovian satellites can be reproduced from the radial temperature gradient of the circumplanetary disk in Canup \& Ward's (2002) model. In the context of this model, Callisto forms on a longer timescale and thereby also avoids differentiation. The rapid inward migration problem is also remedied in the ``gas-starved'' disk model. This model has also been applied to the Saturnian system to successfully explain the incomplete differentiation of Titan \citep{barr10}. %The ``gas-starved'' disk model has been further confirmed in more recent simulations where it forms naturally at the tail-end of the giant planet formation process \citep{coradini10}.
	
A slow infall rate, and resultant low disk mass, is also consistent with current theories of the formation of Saturn's ice-rich ring system \citep{canup10}. \citet{canup10} asserts that Saturn's ring system was formed when a differentiated, Titan-sized satellite migrated to within the Roche limit and its icy mantle was tidally disrupted. The migration of such a satellite requires that some amount of remnant gas still be present in the circum-Saturnian nebula. An upper limit to the amount of gas present is set by the requirement that the ring system not be removed before the circumplanetary nebula completely disperses.  

Recently, \citet{ward10} have developed a comprehensive model that follows the formation and evolution of a giant planet, and subsequent circumplanetary nebula, from cloud collapse through the contraction phase. Their model is composed of three elements: 1) an inflow model describing the properties of the in-flowing material from the circumstellar nebula, 2) a quasi steady-state disk model, and 3) a planet growth and contraction model. It is their quasi steady-state viscous disk model that most concerns us. 

The quasi steady-state viscous disk model requires an in-plane flux as well as a mass loss mechanism at the outer disk boundary, $r_{\rm d}$ \citep{ward10}. Despite the comprehensive nature of their model, \citet{ward10} have yet to identify an appropriate mass loss mechanism and in their own words, ``the outer edge of the disk, $r_{\rm d}$, is not well defined other than it be much further out than the centrifugal radius''. Given a moderate external FUV flux from either the central star or nearby high-mass stars, photoevaporation could provide a natural mechanism for both mass loss and truncation at the outer edge of circumplanetary disks.

\citet{canup02} assume the same circumplanetary disk evolution for both Jupiter and Saturn and that the dichotomy present in their satellite systems is explained by the stochastic timing of formation/migration and the depletion of the solar nebula \citep{canup06}. In contrast, \citet{sasaki10} surmise that the difference in the satellite systems of these two planets is the result of very different disk evolution scenarios. Numerical simulations show that Jupiter is large enough to open a complete gap in the solar nebula cutting off infall onto the Jovian subnebula on a $10^2-10^4$ year timescale \citep{sasaki10}. In contrast, Saturn is too small to have opened a complete gap in the solar nebula and infall onto the Saturnian subnebula would have halted on the $\sim 10^6$ year timescale for the dissipation of the solar nebula as a whole. Although the assumption that Jupiter opened a gap and Saturn did not depends on the assumed viscosity and scale height of the protoplanetary disk, the critical mass for opening a gap is larger in the outer regions of the disk. It is a reasonable assumption using standard disk parameters, and further supported by the current masses of Jupiter and Saturn. For these reasons, \citet{sasaki10} assume in their models that Jupiter opened a gap in the solar nebula and Saturn did not.

\citet{sasaki10} have recently published results from a suite of simulations in which the growth and dynamical evolution of proto-satellite embryos was modeled. They seek to explain why the Jovian regular satellite system consists of four nearly equal mass satellites whereas the Saturnian system contains one large satellite. They propose that the Jovian satellite system may have been ``frozen'' in place when Jupiter grew sufficiently large that a gap was opened in the solar nebula. In the Saturnian system, where only an incomplete gap may have formed, the slower shutoff timescale for material infalling from the solar nebula would have allowed its satellites to continue to dynamically evolve. The typical end result is significant depletion of solids in the inner disk and the retention of a single large satellite in the outer disk that is similar to Titan. This outcome may also be a natural result of the inside-out clearing of the solar nebula due to photoevaporation by the Sun (see discussion in Section \ref{sec:discussion} below).

\citet{sasaki10} are able to produce four or five similarly sized satellites in the Jovian system in $80\%$ of their runs. Whereas, in the Saturnian system only one large satellite remains in $70\%$ of their runs. Their models however rely on a rapid dispersal mechanism for circumplanetary gas once the gap in the solar nebula has been opened. Photoevaporation could provide just such a mechanism as well as help to determine a natural outer disk boundary. As with \citet{ward10}, the models of \citet{sasaki10} rely on an ad hoc outer disk boundary. 

Recent simulations have investigated the tidal truncation of circumplanetary disks \citep{martin10}. These tidal truncation simulations produce disks that are truncated at a radius that occurs at $\sim 0.4\ r_{\rm H}$, where $r_{\rm H}$ is the planet Hill radius. 
\begin{equation}
\label{eqn:hill}
r_{\rm H} = a \sqrt[3]{\frac{M_{\rm p}}{3 M_{\odot}}},
\end{equation}
where $a$ and $M_{\rm p}$ are the planet's semi-major axis and mass and $M_{\odot}$ is the mass of the Sun. This outer disk radius is, however, too large to explain the compact configuration of the regular satellite systems of Jupiter and Saturn which extend to less than $0.06\ r_{\rm H}$.

The primary objective of the present paper is to show how a nominal FUV flux can photoevaporate the outer portions of circumplanetary disks. \citet{adams04} demonstrate how photoevaporation creates a subsonic outflow of gas in the disk that is well inside the gravitational radius, $r_{\rm g}$, where the thermal velocity of the hot disk atmosphere equals the planet's escape velocity.  We apply the \citet{adams04} photoevaporation model to circumplanetary disks and find that the disks are truncated well inside the gravitational radius, $r_{\rm g}$. It is important to note that photoevaporation only needs to remove gas from the planet's Hill sphere in order to truncate the circumplanetary disk; it does not need to remove the gas from the solar system.

The difference in solar flux, due to the difference in semi-major axes, may further account for the different evolutionary histories of Jupiter and Saturn. Saturn's greater distance from the Sun would cause the incident solar flux to be a factor of $(a_{\rm J}/a_{\rm S})^2$ less at Saturn than at Jupiter, where $a_{\rm J}$ and $a_{\rm S}$ are the semi-major axes of Jupiter and Saturn respectively. Even in the compact configuration proposed in the Nice model, the difference in incident flux would be $(5.45\ {\rm AU}/8.18\ {\rm AU})^2 = 0.44$, implying that the incident flux would have been more than twice as strong at Jupiter's location than Saturn's \citep{tsiganis05,morbidelli05}. The greater amount of incident flux at Jupiter would have caused there to be a greater amount of photoevaporative mass loss in the Jovian system. Furthermore, the increased rate of photoevaporation would have caused the Jovian subnebula to be more drastically truncated than the Saturnian subnebula given the same solar luminosity.

%Given a nominal FUV flux, photoevaporation can easily truncate the Jovian subnebula to $50$ Jupiter radii and the Saturnian subnebula to $18$ Saturnian radii. In the Saturnian system this is well within the orbit of Titan and, although in the Jovian system this is nearly twice the orbital radius of Callisto, the disk structure at the locations of the Galilean satellites will certainly be effected by the truncated disk. As stated above, amount the FUV flux at Jupiter's location would be more than twice as large as the incident flux at Saturn's location which would place the truncation radius at the location of Callisto. It is important to note that photoevaporation only needs to remove gas from the planet's Hill sphere in order to truncate the protosatellite disk; it does not need to remove the gas from solar orbit.

We describe our disk model as well as our models for infall from the solar nebula and photoevaporation in Section \ref{sec:disk}. The results of our simulations are presented in Section \ref{sec:results}, with the results from our steady-state disk models presented in Section \ref{subsec:steady} and the results from our time-dependent, decaying disk model in Section \ref{subsec:decay}. In Section \ref{sec:results} we present a summary and discussion of our results.

%%%%%%%%%%%%%%%%%%%%%%%%%%%%%%%%%%%%%%%%%%%%%%%%%%%%%%%%%%%%%%%%%%%%%%%%%%%%%%%%%%%%%%%%%%%%%%%%%%%%%%

\section{Disk Model}
\label{sec:disk}

Our disk model is an extension of an earlier photoevaporative viscous disk model that was previously applied to the evolution of the solar nebula \citep{mitchell10}. Besides its application to circumplanetary nebulae, the current model differs from the previous model in that it includes infall from the solar nebula. The details of the infall will be discussed in Section \ref{subsec:infall}. It is this infall that allows for steady-state solutions.

Our model uses the common $\alpha$-viscosity prescription with a viscosity that is proportional to $r$. 
\begin{equation}
\label{eqn:viscosity}
\nu = \nu_0 (r/R_0)
\end{equation}
where $\nu_0$ and $R_0$ are scalings for viscosity and radius. This assumption has been used in the past by many authors  \citep{clarke07,hartmann98b}. The linear dependence of viscosity on radius in our model implies that the temperature profile is proportional to $r^{-1/2}$. The midplane disk temperature, $T_{\rm disk}$, is used in order to determine the viscosity constant $\nu_0$.
\begin{equation}
\nu_0 = \alpha \sqrt{\frac{k T_{\rm disk}}{\muh}}
\end{equation}
where $k$ is the Boltzmann constant and $\muh$ is the mean molecular weight. We have scaled the midplane temperature profiles in our models such that in the Jovian system the temperature has been set to $\approx 250\ {\rm K}$ at $10\ R_{\rm J}$. This is consistent with the slow-inflow, low-opacity circum-Jovian accretion disk model investigated by \citep{canup02}. For the Saturnian system, the radial temperature profiles in our models have been set to $\approx 100\ {\rm K}$ at $20\ R_{\rm S}$, the approximate location of Titan.

Most models of circumplanetary disks employ a value of $\alpha$ that is generally smaller than that used in circumstellar disk models \citep{canup09}. The simulations presented in this work consider $\alpha$ values in the range of $10^{-2}-10^{-4}$, with our reference models having a value of $\alpha = 10^{-3}$.

Instead of using $r$ and $\Sigma$; the radius and mass surface density of the gas, it is useful to describe the system in terms the variables $h$ and $g$; the specific angular momentum and torque. This is a good choice of variables to make when the viscosity is proportional to $r$ as it allows us to transform the viscous disk equation into a simple, linear differential equation with a constant coefficient \citep{hartmann98a}. 
\begin{equation}
\label{eqn:h}
h = r^2 \Omega = (G M_{\rm p} r)^{1/2}
\end{equation}

\begin{equation}
\label{eqn:g}
g = -2 \pi r \Sigma \nu r^2 \frac{d \Omega}{d r} = 3 \pi \nu h \Sigma
\end{equation}                                  
where $G$ is the gravitational constant and $\Omega$ is the Keplerian orbital velocity. The relationship between $g$ and $\Sigma$ in Eqn. (\ref{eqn:g}) implies that the mass surface density is inversely proportional to the viscosity. Thus, $\Sigma\propto \alpha^{-1}$.

By substituting in the functional form of viscosity, the continuity equation for radial mass transport can be written as
\begin{equation}
\label{eqn:viscous}
\frac{\partial g}{\partial t} = \frac{3}{4} \frac{\nu_0 G M_{\rm p}}{R_0} \frac{\partial^2 g}{\partial h^2},
\end{equation}
where $M_{\rm p}$ is the mass of the planet.

As we are interested in the addition of material to the disk, a source term must be added onto the right-hand side of Eqn. (\ref{eqn:viscous}).
\begin{equation}
\label{eqn:visc_source}
\frac{\partial g}{\partial t} = \frac{3}{4} \frac{\nu_0 G M_{\rm p}}{R_0} \frac{\partial^2 g}{\partial h^2} + \frac{3 \pi \nu_0}{G M_{\rm p} R_0} h^3 \dot{\Sigma}(h,t),
\end{equation}
where $\dot{\Sigma}(h,t)$ is the infall rate in [${\rm g\ cm^{-2} s^{-1}}$] as a function of specific angular momentum (radius) and time.

%%%%%%%%%%%%%%%%%%%%%%%%%%%%%%%%%%%%%%%%%%%%%%%

\subsection{Infall}
\label{subsec:infall}

The partial differentiation of Callisto implies a long formation timescale and was one factor that led \citet{canup02} to develop their ``gas-starved'' disk model. They assert that the $0.02$ Jupiter masses required for formation the Galilean satellites, using a minimum mass solar nebula approach, need not be present all at once. This material may have slowly flowed through the system over an extended period of time. In this scenario, the agglomeration of solids into satellites is analogous to the build-up of mineral deposits in plumbing over time. They introduce a timescale for the addition of mass to the system.

\begin{equation}
\tau_{\rm G} = \frac{M_{\rm p}}{\dot M_{\rm p}},
\end{equation}
In all simulations presented here, we use a gas accretion timescale of $\tau_{\rm G} = 5\times10^6$ yr. This is equivalent to the mass accretion rate of $2\times 10^{-7} M_{\rm P}\ {\rm yr}^{-1}$ which was found by \citet{canup02} to be consistent with the conditions needed to form the Galilean satellites. This mass accretion rate is also consistent with gap accretion simulations \citep{bryden99}.

The location of the outer boundary is not only governed by photoevaporation, but by viscous spreading as well. It is therefore important to consider the location and amount of infalling material from the solar nebula. If the majority of the infalling material falls onto the outer regions of the disk, because of its angular momentum content, it will affect the behavior of the outer boundary. It is therefore important to parameterize the location and amount of infalling material, and not just treat it as a constant throughout the disk as do \citet{canup02,canup06}.

\citet{machida08} ran high resolution three-dimensional simulations of the accretion of angular momentum onto a protosatellite system. Using their results, we can estimate the centrifugal radius, $r_{\rm c}$, the location at which the bulk of the infalling gas lands on the circumplanetary nebula. We estimate that, for a Jupiter mass protoplanet, the infalling material has an average specific angular momentum consistent with a Keplerian orbit at $r_{\rm c} \approx 25\ R_{\rm J}$. We use the angular momentum contained within the infalling material to constrain the radii where infall occurs, so that the added disk mass has the same angular momentum than the local circular orbit where it is added. Were this not the case, and the infalling material contained less angular momentum as the local circular orbit where it was added, it would not be rotationally supported and would rapidly fall inward and be redistributed at smaller radii corresponding to its angular momentum content. A very recent, high-resolution simulation of mass accreted from the solar nebula onto circumplanetary disks indicates that infalling mass does, if fact, intercept the disk at radii ($\sim 50 R_{\rm J}$) greater than that which corresponds to its angular momentum content and is redistributed inward \citep{tanigawa11}. While these results are too new to have been included in this investigation, they will certainly be taken into consideration in future investigations. 

We have adopted the following functional form, which peaks at $25\ r_{\rm p}$, for the infalling mass onto our protosatellite nebulae.

\begin{equation}
\label{eqn:infall}
\dot{\Sigma}(r,t) = -\biggl[\biggl(\frac{r}{r_{\rm p}}\biggr) - 28\biggr] {\rm exp}\Biggl[\frac{(\frac{r}{r_{\rm p}}) - 28}{3}\Biggr]\ {\rm g\ cm^{-2} s^{-1}},
\end{equation}
where $r_{\rm p}$ is the planet's current equatorial radius. The prescription for the infalling material has roughly $90\%$ of the infalling mass accreting onto the circumplanetary disk at radii between $16\ R_{\rm p} < r < 28\ R_{\rm p}$ and peaks at $25\ R_{\rm p}$. This prescription is consistent with the simulations of \citet{machida08} which show that the infalling material has some angular momentum distribution around the peak which corresponds to a centrifugal radius of, $r_{\rm c} = 25\ R_{\rm p}$. Our prescription for infall is also consistent with the previous work of \citet{canup02,canup06} in which the infall is limited to an inner region of the disk. In our model there is no accretion of material onto the circumplanetary disks from the solar nebula external to $r = 28\ R_{\rm p}$.

Once the gas giant planets grow large enough, they begin to open gaps in the solar nebula as a result of resonant interactions with the disk. The gap-clearing timescale can be estimated by assuming that it would occur on the viscous timescale to spread across the scale height in the protoplanetary disk \citep{sasaki10}. 
\begin{equation}
\tau_{\rm gap} \sim \frac{H_{\rm PD}^2}{\nu} \sim (10^{-3}-10^{-4})\times (1-10)\ {\rm Myr}
\end{equation}
The actual timescale for a growing planet to open a gap is likely longer than this estimate, but it must be substantially shorter than the planet's accretion timescale in order to limit the final mass of the planet. A reasonable estimate for the gap opening timescale would be to assume a median value of $\tau_{\rm gap} \approx 2.5\times 10^3$ yr.

We assume that the infall rate decays exponentially over some timescale, $\tau_{\rm off}$.
\begin{equation}
\label{eqn:infall_decay}
\dot{\Sigma}(t) = \dot{\Sigma}\ {\rm exp}\biggl[\frac{t}{\tau_{\rm off}}\biggr]
\end{equation}
This assumed form for the infall decay rate is also made by other authors in similar investigations \citep{canup10}. We further assume that the infall decay time is of the same order as the gap opening timescale, $\tau_{\rm off} = \tau_{\rm gap} = 2.5\times 10^3$ yr.

Although it seems likely that gas continues to accrete through the gap in a circumstellar disk opened by a growing protoplanet, it is uncertain if this gas is able to accrete either onto the protoplanet or even onto a circumplanetary disk. We have chosen the exponential decay of infalling material shown in Eqn. \ref{eqn:infall_decay} because it is simple and easy to understand. Also, it allows a direct comparison with \citet{sasaki10}, which use the same prescription for the exponential decay and short infall decay timescales. Furthermore, whether it be via gap opening or the global depletion of the solar nebula, it is certain that the infall from the solar nebula should halt at some point in time. The only uncertainty is the timescale over which the infall wanes.

%%%%%%%%%%%%%%%%%%%%%%%%%%%%%%%%%%%%%%%%%%%%

\subsection{Photoevaporation}
\label{subsec:photoevaporation}

Recent observations of $56$ weak-lined and classical T Tauri stars was used to determine the FUV emission of young $1-10$ Myr stars and whether the observed flux is consistent with that required by models of photoevaporative dispersal of circumstellar disks \citep{ingleby11}. These authors concluded that radiation fields sufficiently strong for the removal of gas are present during the disk dispersal phase. Although not considered in this study, x-ray flux has been shown to enhance FUV-driven photoevaporation \citep{gorti09}. \citet{ingleby11} also investigated the x-ray emission of $1-10$ Myr stars and found that the x-ray flux remains high, and constant, throughout the duration of the dispersal phase.

Clearly, the young Sun would have exposed the circumplanetary nebulae of Jupiter and Saturn to FUV radiation. Although it is unclear at this time how much FUV from the Sun is able to reach these circumplanetary nebulae, one must remember that the Sun most likely formed in a cluster of $10^3-10^4$ stars which could have contributed to the FUV field in which these nebulae were embedded \citep{fatuzzo08,adams10}. Whatever its source, the FUV radiation would have heated the periphery of the circumplanetary disks. Gas heated to sufficient temperatures would then have become unbound from these disks. The gravitational radius, $r_{\rm g}$, is defined as the radius at which the sound speed of the heated gas equals the escape speed from the system,
\begin{equation}
\label{eqn:radius}
r_{\rm g} = \frac{G M_{\rm p} \muh}{kT},
\end{equation}
where $k$ is the Boltzmann constant and $T$ is the temperature of the super-heated atmosphere, or what we will refer to as the envelope temperature, $T_{\rm env}$. Gas beyond the gravitational radius will escape from the system. The gravitational radius is the canonical radius beyond which gas heated to a temperature $T_{\rm env}$ will escape from the disk. In actuality, gas can escape from the disk at radii substantially smaller than $r_{\rm g}$.

The heating of the disk's surface by FUV radiation and the resultant outflow are complicated processes, but it is useful to employ a simplified model with an isothermal, heated atmosphere. Consider a disk irradiated and heated by external FUV radiation from either the early Sun or the fellow members of its birth cluster. Depending on the strength of the FUV flux, the heated gas will reach temperatures in the range $100\ {\rm K} < T< 3000\ {\rm K}$ \citep{adams04}. As the gas heats, it expands generating a neutral outflow. The expanding outflow begins subsonically but becomes supersonic by the time it reaches the gravitational radius. This outflow is generally isotropic, but the majority of mass loss is dominated by mass loss from the outer edge of the disk. The isotropic, neutral outflow serves to shield the disk from EUV radiation that would ionize the disk and heat it to $\approx 10,000$ K. See \citet{johnstone98,adams04} for more thorough discussions.

\citet{adams04}  studied the previously unexplored subcritical regime, where the outer radius of the disk, $r_{\rm d}$, is smaller than the gravitational radius. Using results from their photodissociation region code, they developed analytical models for the photoevaporative mass loss rates for cases in which $r_{\rm d}$, the location of the outer edge of the disk, is both inside and outside the gravitational radius. These models are characterized by a single temperature, $T_{\rm env}$, which determines a unique sound speed, $a_{\rm s}$, for the isothermal atmosphere.
\begin{mathletters}
\begin{eqnarray}
\lefteqn{\hspace{-.5in} \dot{M}  =  C_0 N_{\rm C} \muh a_{\rm s} r_{\rm g} \Bigl( \frac{r_{\rm g}}{r_{\rm d}} \Bigr) {\rm exp} \Bigl(- \frac{r_{\rm g}}{2r_{\rm d}}\Bigr)} \nonumber \\ 
& & \hspace{1.0in} r_{\rm d} < \beta r_{\rm g} \label{eqn:massloss_a} \\  
\lefteqn{\hspace{-.5in}\dot{M}  =  4 \pi {\mathcal F} \muh \sigma^{-1}_{\rm FUV} a_{\rm s} r_{\rm d}} \nonumber \\ 
& & \hspace{1.0in} r_{\rm d} > \beta r_{\rm g} \label{eqn:massloss_b} 
\end{eqnarray}
\end{mathletters}
The first equation is for subcritical disks, and the second equation is for supercritical disks, where $a_{\rm s}$ is the sound speed in the heated atmosphere and $C_0$ is a constant of order unity. $N_{\rm C}$ is the critical surface density of the flow and $\sigma_{\rm FUV}$ is the cross section for dust grains interacting with FUV radiation. The dust optical depth is given by $\tau_{\rm FUV} = \sigma_{\rm FUV} \cdot N_{\rm H}$. For an optical depth of order unity, $\sigma^{-1}_{\rm FUV}\approx N_{\rm H}$, where $N_{\rm H}$ is evaluated at the critical density $N_{\rm C} \sim 10^{21}\ {\rm cm}^{-2}$. The factor $C_0$ is a constant of order unity used by Adams et al. (2004) to match their numerical and analytical solutions. It is used in our model to match the mass loss rates for sub- and supercritical disks at a radius of $r_{\rm d}/r_{\rm g} = 0.25$ and is adopted as our value of $\beta$. This is necessary because the mass loss rates are sensitive functions of strength of the FUV radiation field as well as the assumed matching point. 

We have matched the subcritical solution onto the supercritical solution, because the supercritical solution is better understood and well constrained. The factor ${\mathcal F}$, in the second equation, is the fraction of the solid angle subtended by the flow and is $\sim 1$ because the flow from the disk surface and edge merge at a radius between $r_{\rm d}$ and $2\cdot r_{\rm d}$; creating a nearly spherically symmetric outflow \citep{adams04}.

The differential equation governing the location of the outer boundary was derived using mass conservation at the outer boundary \citep{mitchell10}. 

\begin{equation}
\label{eqn:boundary_mass}
\dot M_{\rm boundary \atop motion} = \dot M_{\rm viscous \atop spreading} - \dot M_{\rm photo- \atop evaporation}
\end{equation}

The mass flux due to viscous processes is simply $- \partial g/\partial h$ and the mass flux due to photoevaporation is taken to be Eqn. (\ref{eqn:massloss_a}) or Eqn. (\ref{eqn:massloss_b}) depending on the location of $r_{\rm d}$. The mass flux due to the motion of the boundary is $2 \pi r_{\rm d} \Sigma_d (dr_{\rm d}/dt)$. Here, and throughout the paper, the subscript ``d'' indicates that these quantities are to be evaluated at the outer disk edge. The equations governing the location of the outer boundary were derived by substituting these expressions into Eqn. (\ref{eqn:boundary_mass}). Depending on the location of the outer boundary, we will either be in the subcritical regime,

\begin{equation}
\label{eqn:subcritical}
\frac{d r_{\rm d}^2}{d t} = - \frac{1}{\sqrt{2 \pi} N_{\rm c} \muh}\frac{\partial g}{\partial h_{\rm d}} - \frac{C_0 a_{\rm s} r_{\rm g}}{\sqrt{2 \pi}} \biggl( \frac{r_{\rm g}}{r_{\rm d}} \biggr) {\rm exp}\biggl[-\frac{r_{\rm g}}{r_{\rm d}}\biggr] 
\end{equation}

or the supercritical regime,

\begin{equation}
\label{eqn:supercritical}
\frac{d r_{\rm d}^2}{d t} = -\frac{1}{\sqrt{2 \pi} N_{\rm c} \muh} \frac{\partial g}{\partial h_{\rm d}} - \sqrt{2 \pi}\ a_{\rm s} r_{\rm d},
\end{equation}

At each time step, depending on whether we are in the sub- or supercritical regime, either Eqn. (\ref{eqn:subcritical}) or Eqn. (\ref{eqn:supercritical}) must first be solved to determine the new boundary location. Then the viscous evolution, Eqn. (\ref{eqn:visc_source}) is solved. All of the simulations presented here were performed on a Mac G5 PowerPC. They were evolved on a grid of 201 points evenly spaced in non-dimensional specific angular momentum. 

Our initial conditions are that of the similarity solutions of \citet{lynden-bell74}. 
\begin{equation}
\label{eqn:initial}
\Sigma_{\rm init} = \frac{M_0}{2 \pi R_1 r} {\rm exp}\biggr[-\frac{r}{R_1}\biggr]
\end{equation}
where $R_1$ is the initial disk scaling radius and $M_0$ as the initial disk mass.

As with most previous disk models, we employ a zero torque inner boundary condition. This allows for accretion from the disk onto the planet. Our outer boundary condition is set by the torque exerted on the disk by the out-flowing material that is being photoevaporated at the disk's edge. The torque exerted on the outer edge of the disk can be expressed as 
\begin{equation}
{g}_{\rm d} = 3\sqrt{2\pi}N_{\rm c}\muh\nu_0\biggl(\frac{r_{\rm d}}{R_0}\biggr)h_{\rm d}.
\end{equation}

It must be noted that these two boundary conditions are used to solve the equation governing the temporal evolution of the mass surface density. We have, in effect, a second outer boundary condition that governs the temporal evolution of the location of the outer boundary. This condition is set by a differential equation that was derived using mass conservation at the outer boundary (Eqn.'s (\ref{eqn:subcritical}) \& (\ref{eqn:supercritical})).

%%%%%%%%%%%%%%%%%%%%%%%%%%%%%%%%%%%%%%%%%%%%%%%

\subsection{Variable Space Grid Method}
\label{subsec:vsg} 

A well-posed problem in the material sciences, called the Stefan Problem, deals with propagating phase changes typically considered in the context of melting/freezing and heat ablation \citep{ozisik80}. We have developed a one-dimensional model of an astrophysical disk that includes viscous diffusion and photoevaporation at the outer boundary using numerical techniques developed to solve the Stefan Problem \citep{mitchell10}. By adapting the Stefan problem to astrophysical disks, our one-dimensional numerical model self-consistently tracks the location of the outer boundary. This is a novel approach to modeling astrophysical disks.

To solve the Stefan problem, \citet{kutluay97} adopt a numerical method with a variable space grid (VSG) first proposed by \citet{murray59}. The VSG method employs a fixed number of grid points with a variable grid size at each time step. This method involves solving two coupled differential equations at each time step, one for the location of the outer boundary and one for the diffusive evolution of the disk. Once the location of the outer boundary is found the abscissa is rescaled and the diffusive evolution calculated. Please refer to \citet{mitchell10} for further details.

%%%%%%%%%%%%%%%%%%%%%%%%%%%%%%%%%%%%%%%%%%%%%%%%%%%%%%%%%%%%%%%%%%%%%%%%%%%%%%%%%%%%%%%%%%%%%%%%%%%

\section{Results}
\label{sec:results}

\subsection{Steady-State Disks}
\label{subsec:steady}

We present a number of simulations of steady-state circumplanetary disks about Jupiter and Saturn. These simulations produce steady-state solutions by coupling viscous evolution, photoevaporation and infall from the solar nebula. A goal of this work is to investigate the range of possible truncation radii and disk masses based on the amount of FUV flux and the strength of the viscosity parameter, $\alpha$. 

These simulations all agree well with the low surface densities found by \citet{canup02}. As in the simulations of \citep{machida08}, we see surface density enhancements at roughly $25\ r_{\rm p}$, as a result of infall from the solar nebula. Such enhancements are not seen in recent, high-resolution simulations in which the infall occurs over a wide range of radii \citep{tanigawa11}. These enhancements likely arise from our choice of infall shape (Eqn. (\ref{eqn:infall})) and the fact that our viscosity depends only on radius and not on the local conditions in the circumplanetary disk. If such density enhancements are real, they may have implications for satellite growth. We will further explore the significance of these density enhancements in Section \ref{sec:discussion}.

\begin{figure}[ht]
\begin{center}
\includegraphics[width=2.25in]{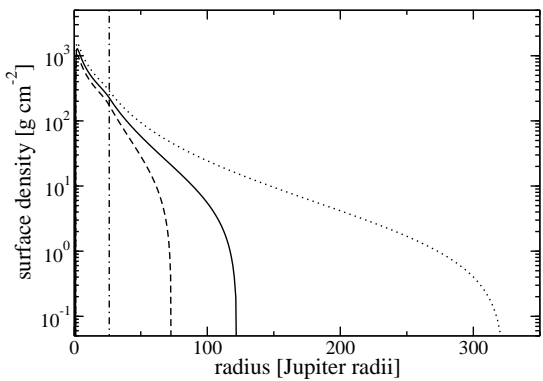}
\caption[jup_temp]
	{
	\label{fig:jup_temp}
Steady-state radial mass surface density of Jupiter's circumplanetary disk. The three curves are for three different values of the isothermal, heated envelope. The solid line is of our reference model with $T_{\rm env} = 600\ {\rm K}$, whereas the dotted and dashed curves are for simulations with $T_{\rm env} = 100\ {\rm K}$ and $T_{\rm env} = 3000\ {\rm K}$ respectively. The current location of Callisto has been marked with a dot-dashed line.
	}
\end{center}
\end{figure}

The UV flux can produce a wide range of envelope temperatures ($100\ {\rm K}-3000\ {\rm K}$) depending on the magnitude of the flux \citep{adams04,mitchell10}. As stated in Section \ref{subsec:photoevaporation}, recent observations confirm that the FUV emission of young, solar-type stars is sufficient for photoevaporation \citep{ingleby11}. Figure \ref{fig:jup_temp} shows the radial mass surface density from three simulations in which the envelope temperature has been varied. The solid line is of our reference model with $T_{\rm env} = 600\ {\rm K}$, whereas the dotted and dashed curves are for simulations with $T_{\rm env} = 100\ {\rm K}$ and $T_{\rm env} = 3000\ {\rm K}$ respectively. These simulation were all run with $\alpha = 10^{-3}$. One significant feature of these simulations is the enhancement in mass at $r\approx 25\ R_{\rm J}$ mentioned above. The truncation radii of these simulations ranges from $73\ R_{\rm J}$ in our high-temperature simulation to $324\ R_{\rm J}$ in our low temperature simulation. In terms of Hill's radius, these disk radii range from $0.098-0.44\ r_{\rm H}$ with our fiducial model's outer radius at $0.17\ r_{\rm H}$.

The higher envelope temperature causes more erosion at the outer boundary and therefore causes the steady-state disk to have a much smaller truncation radius. Despite having a much smaller truncation radius, the three runs presented in Figure \ref{fig:jup_temp} all have disk masses that lie in a narrow range from $3.3\times 10^{-5}-1.2\times 10^{-4}\ M_{\rm J}$.

\begin{figure}[ht]
\begin{center}
\includegraphics[width=2.25in]{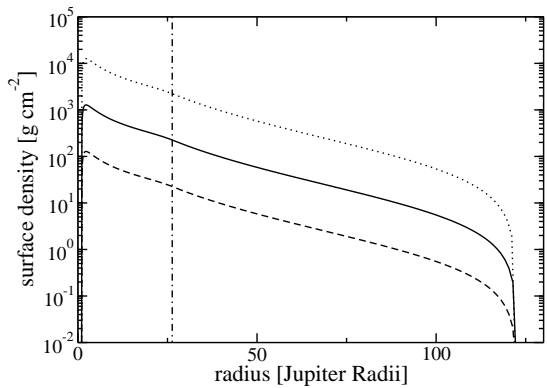}
\caption[jup_visc]
	{
	\label{fig:jup_visc}
Steady-state radial mass surface density of Jupiter's circumplanetary disk. The three curves are for three different values of the viscosity parameter, $\alpha$. The solid line is of our reference model with $\alpha = 10^{-3}$, whereas the dotted and dashed curves are for simulations with $\alpha = 10^{-4}$ and $\alpha = 10^{-2}$ respectively. The current location of Callisto has been marked with a dot-dashed line.
	}
\end{center}
\end{figure}

We have also run a suite of simulations in which the viscosity parameter, $\alpha$, has been varied. We have investigated $\alpha$ in the range of $10^{-4}-10^{-2}$. Simulations of steady-state circum-Jovian disks are presented in Figure \ref{fig:jup_visc}. The solid line is of our reference model with $\alpha = 10^{-3}$, whereas the dotted and dashed curves are for simulations with $\alpha = 10^{-4}$ and $\alpha = 10^{-2}$ respectively. These simulation were all run with $T_{\rm env} = 600$ K.  A striking difference between the varied viscosity runs and those for a varied envelope temperature, presented in Figure \ref{fig:jup_temp}, is that the outer disk radius is independent of the strength of the viscosity. It depends only on the mass loss rate at the outer boundary which, in our models, is controlled solely by the envelope temperature.

Another difference between the temperature and viscosity runs is in the masses of the steady-state disks. The disk masses in the runs presented in Figure \ref{fig:jup_visc} each differ by an order of magnitude and lie in the range of $5.5\times 10^{-6}-5.4\times 10^{-4}\ M_{\rm J}$. This suggests that the steady-state disk mass is dominated by the strength of the viscosity and not by the amount of photoevaporative mass loss.

\begin{figure}[ht]
\begin{center}
\includegraphics[width=2.25in]{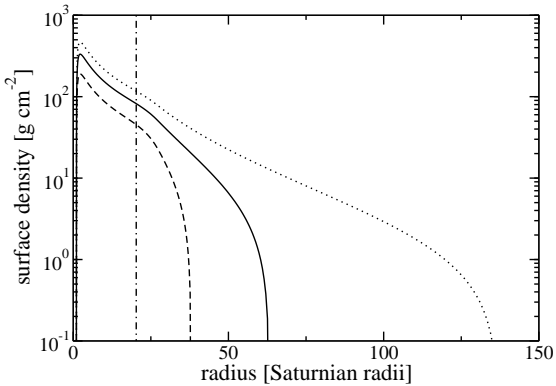}
\caption[sat_temp]
	{
	\label{fig:sat_temp}
Steady-state radial mass surface density of Saturn's circumplanetary disk. The three curves are for three different values of the isothermal, heated envelope. The solid line is of our reference model with $T_{\rm env} = 600\ {\rm K}$, whereas the dotted and dashed curves are for simulations with $T_{\rm env} = 100\ {\rm K}$ and $T_{\rm env} = 3000\ {\rm K}$ respectively. The current location of Titan has been marked with a dot-dashed line.
	}
\end{center}
\end{figure}

\begin{figure}[ht]
\begin{center}
\includegraphics[width=2.25in]{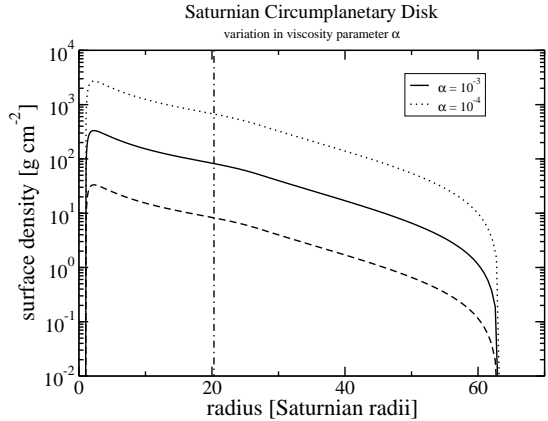}
\caption[sat_visc]
	{
	\label{fig:sat_visc}
Steady-state radial mass surface density of Saturn's circumplanetary disk. The three curves are for three different values of the viscosity parameter, $\alpha$. The solid line is of our reference model with $\alpha = 10^{-3}$, whereas the dotted and dashed curves are for simulations with $\alpha = 10^{-4}$ and $\alpha = 10^{-2}$ respectively. The current location of Titan has been marked with a dot-dashed line.
	}
\end{center}
\end{figure}

We have conducted a similar suite of simulations for the Saturnian system as for the Jovian system. Simulations in which we have varied the envelope temperature are presented in Figure \ref{fig:sat_temp} and those in which we have varied the strength of the viscosity are presented in Figure \ref{fig:sat_visc}. The same range of parameter space has been explored in these runs as was done for the Jovian system. These data are presented with the same line scheme as those presented in Figures \ref{fig:jup_temp} and \ref{fig:jup_visc}. 

Due to its location farther out in the solar nebula, the conditions during formation were probably somewhat different at Saturn than at Jupiter. The inside-out clearing of the solar nebula by photoevaporation described by \citet{gorti09} would have caused the solar FUV flux to vary significantly in space and time. It is however almost certain that both Jupiter and Saturn experienced some migration during the formation process \citep{ward97,ward98}. The direction and rate of migration are highly uncertain and we cannot say for sure where either of these systems formed \citep{paardekooper06,crida07,walsh11}. Furthermore, we feel that our range of parameter space is appropriate for both the Jovian and Saturnian systems and exploring the same range in the two systems is best for comparison.

Once again, one can see in the simulations of the Saturnian system that the total disk mass is dependent on the strength of the viscosity whereas the radial extent of the modeled disks is dominated by the envelope temperature. The disk masses for each disk presented in Figure \ref{fig:sat_visc} also vary by an order of magnitude, just like those presented in Figure \ref{fig:jup_visc} and lie in the range of $2.3\times 10^{-6}-1.9\times 10^{-4}\ M_{\rm S}$. The outer disk radii of the disk models presented in Figure \ref{fig:sat_temp} span a range from $38-137\ R_{\rm S}$, which in terms of the planet's Hill radius is $0.035-0.13\ r_{\rm H}$. Despite the wide range of outer disk radii, the simulations presented in Figure \ref{fig:sat_temp} have disk masses in a narrow range from $1.0\times 10^{-5}-5.1\times 10^{-5}\ M_{\rm S}$.

The time-dependent, viscous evolution of our models allow us to accurately track the transfer of mass throughout our disks. The transfer of mass is important for the formation of satellites. The rate of mass transfer through the disk, $\dot{M}$ can be calculated at any radius by simply taking the derivative if the torque with respect to the specific angular momentum. 
\begin{equation}
\label{eqn:mass_transfer}
\dot{M} = -\frac{\partial g}{\partial h}
\end{equation}

Disk models are often characterized by the slope of the radial mass surface density using a power law of the form
\begin{equation}
\label{eqn:power_law}
\Sigma(r) \propto r^{-q}
\end{equation}
The slope of the radial surface mass density profiles of the models presented in Figures \ref{fig:jup_visc} and \ref{fig:sat_visc} are roughly power laws with $q = -1.3$. This slope is steeper than the slopes used in the disk models of \citet{canup02,sasaki10}, who both assume $q = -3/4$. However, it is in accordance with our assumed radial temperature dependence 0f $r^{-1/2}$, which predicts $q = -1$. The fact that $q$ is slightly steeper than $q = -1$ in our models can be attributed to the truncation of the outer boundary by photoevaporation. It must be kept in mind that all of the slopes mentioned here are ad hoc and have been assumed by the various authors. It is unclear how steep these disks would actually be given a more comprehensive model with realistic viscosity that depends on the local conditions in the disk.

\begin{figure}[ht]
\begin{center}
\includegraphics[width=2.25in]{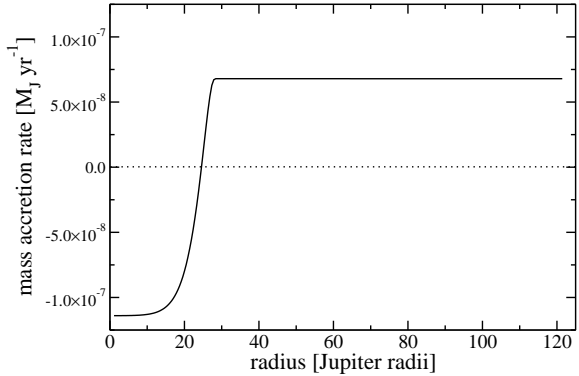}
\caption[mass_transfer]
	{
	\label{fig:mass_transfer}
Mass transfer rate in a steady-state circum-Jovian disk as a function of radius. This analysis was conducted on our reference model, with an envelope temperature of $600$ K and a viscosity parameter $\alpha=10^{-3}$. Nearly equal mass is transferred inward and outward in the disk, with slightly more mass accreted onto the Jupiter than is lost through the outer edge due to photoevaporation. The mass accretion rate in the outer region of the disk is characterized by a constant mass flux rate. The region of constant flux begins exterior to $28\ R_{\rm J}$. Interior to $28\ R_{\rm J}$, the mass flux rate is decreases and at $\sim 24\ R_{\rm J}$ it becomes negative. The region inward of $28\ R_{\rm J}$ corresponds to the region over which infall from the solar nebula occurs. Inward of $\sim 24\ R_{\rm J}$ the mass flux rate is negative and is being accreted onto the planet.
	}
\end{center}
\end{figure}

Figure \ref{fig:mass_transfer} shows the mass transfer rate as a function of radius in a steady-state circum-Jovian disk. This analysis was conducted on our reference model, with an envelope temperature of $600$ K and a viscosity parameter $\alpha=10^{-3}$. 
Nearly equal mass is transferred inward and outward in the disk, with slightly more mass accreted onto the planet than is lost through the outer edge due to photoevaporation. Of the $2\times 10^{-7}\ M_{\rm J}\ {\rm yr}^{-1}$ of material being accreted from the solar nebula, $1.25 \times 10^{-7}\ M_{\rm J}\ {\rm yr}^{-1}$ is accreted onto Jupiter and  $0.75 \times 10^{-7}\ M_{\rm J}\ {\rm yr}^{-1}$ is lost through photoevaporation at the disk's outer edge. Assuming a solar dust-to-gas mass ratio of $0.014$, this is sufficient to provide enough mass in solids through the outer regions of the disk to build Callisto over the $10^5$ yr required for it to remain undifferentiated \citep{canup02}. 
\begin{equation}
M_{\rm tot} = f \dot{M} \tau_{\rm acc} = 1.9\ M_{\rm Callisto}
\end{equation}
where $f$ is the assumed dust-to-gas mass ratio of $0.014$ and $\tau_{\rm acc}$ is the accretion timescale of $10^5$ yr.

%%%%%%%%%%%%%%%%%%%%%%%%%%%%%%%%%%%%%%%%%

\subsection{Decaying Infall}
\label{subsec:decay}

\begin{figure}[ht]
\centering
\includegraphics[width=2.25in]{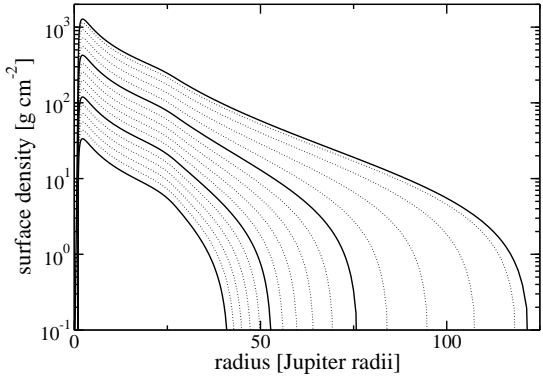}
\caption[decay]
	{
	\label{fig:decay}
Temporal evolution of Jupiter's circumplanetary disk as the infall from the solar nebula exponentially decays. The radial mass surface density is shown at $500$yr increments, with the solid bold lines indicating $t = \tau_{\rm off}, 2\cdot \tau_{\rm off}\ {\rm and}\ 3\cdot \tau_{\rm off}$.
	}
\end{figure}

A secondary goal of this work is to validate photoevaporation as a potential mechanism for the rapid dispersal of circumplanetary nebulae as infall from the solar nebula wanes because of gap opening. We have performed one such simulation that investigates the decrease in infall rate onto the Jovian subnebula over a $2.5\times 10^3$ yr timescale (see Section \ref{subsec:infall}). Figure \ref{fig:decay} shows the temporal evolution of Jupiter's circumplanetary disk as the infall is abated. 

Similarly to \citet{sasaki10}, we have begun this simulation in the context of the ``gas-starved'' disk model of \citet{canup02}. This is done because gap opening would likely occur in the final stage of giant planet accretion. Therefore, the simulation presented in Figure \ref{fig:decay} was begun with the steady-state solution from our Jovian reference model with $\tau_{\rm G} = 5\times 10^6$ yr. This value differs slightly from that used by \citet{sasaki10} of $2\times 10^6$ yr, but as they have shown the distribution and composition of final satellites is insensitive to the exact value of $\tau_{\rm G}$. The radial surface density is shown at $500$ yr increments, with the solid bold lines indicating $t = \tau_{\rm off}, 2\cdot \tau_{\rm off}\ {\rm and}\ 3\cdot \tau_{\rm off}$. The total mass of the circumplanetary disk decreased nearly two orders of magnitude from $5.5\times 10^{-5}\ M_{\rm J}$ to $7.6\times 10^{-7}\ M_{\rm J}$ over the course of this simulation.

This simulation was done using a nominal FUV flux, which corresponds to an envelope temperature of $600$ K. Envelope temperatures can range from $100$ K to $3000$ K \citep{adams04}. This shows that, even for a moderate FUV flux, photoevaporation is able to clear the Jovian subnebula on the very short timescale over which the infall wanes due to gap opening.

%%%%%%%%%%%%%%%%%%%%%%%%%%%%%%%%%%%%%%%%%%%%%%%%%%%%%%%%%%%%%%%%%%%%%%%%%%%%%%%%%%%%%%%%%%%%%%%%%%

\section{Summary and Discussion}
\label{sec:discussion}

We have modeled and analyzed circumplanetary, protosatellite disks with the combined influence of viscous forces and photoevaporation. Our models also include mass infall from the solar nebula allowing for steady-state solutions. These models were developed in the context of the ``gas-starved'' disk models put forth by \citep{canup02}. With these models, we present a new mechanism for the truncation of circumplanetary disks. We model the late stages of giant planet growth, when the accretion rate from the solar nebula is limited. The limited accretion rate may be a result of gap opening, the global depletion of the solar nebula or a combination of both.
 
The regular satellites of Jupiter and Saturn extend to $0.06\ r_{\rm H}$. This value is much smaller than the extent of the circumstellar disk predicted considering the angular momentum content of the accreting gas. By estimating the angular momentum of accreting gas as it travels from the solar nebula, through the Lagrange points, and onto Jupiter, \citet{quillen98} estimated that the truncation radius is $\sim r_{\rm H}/3$. A similar outer disk radius was found in numerical simulations which consider only the angular momentum content of accreting gas \citep{ayliffe09}. The recent, high-resolution simulation of \citet{tanigawa11} indicates that accreting gas intercepts the Jovian circumstellar disk at $\sim r_{\rm H}/15$ and that much of the gas has an angular momentum content that corresponds to even smaller radii than this.

Recently, the radial extent of the regular satellites of Jupiter and Saturn has been explained by the rapid accumulation of solids at the location of the infalling material, $r_{\rm c}$ or by the truncation of the protosatellite disk by solar tides \citep{canup02,martin10}. We invoke photoevaporation to naturally explain the location of regular satellites. Our models produce a large range of outer disk radii depending on our choice of envelope temperature (FUV flux). These outer disk edges range in radii from $0.035\ r_{\rm H}$ to $0.44\ r_{\rm H}$, with a mean value of $0.16\ r_{\rm H}$. Even though our models show a wide range of outer disk edge radii based on envelope temperature, the masses of these disk all lie in a narrow range from $3.3\times 10^{-5}-1.2\times 10^{-4}\ M_{\rm J}$ in the Jovian models and $9.7\times 10^{-6}-5.1\times 10^{-5}\ M_{\rm S}$ in the Saturnian system. This narrow range of disk masses is to be expected because these simulations were all run with the same alpha-viscosity parameter of $10^{-3}$ 

The disks produced in our Saturnian models extend to a much smaller extent of their Hill spheres because Saturn is less than a third as massive as Jupiter and nearly twice as far from the Sun. In relation to their masses, Saturn's Hill sphere is much larger than Jupiter's. Although the size of a planet's Hill radius plays no role in our simulations, similarly sized disks, relative to their host planet's mass, would extend to different fractions of their respective Hill spheres.

Our models show that photoevaporation can truncate circumplanetary disks to radii that are in concordance with the locations of the regular satellites of the giant planets in our solar system. Our reference model produces outer disk radii truncated at $0.057\ r_{\rm H}$ and $0.17\ r_{\rm H}$ for Saturn and Jupiter respectively. These small outer disk radii may provide an obvious explanation for the locations of the regular satellites of Jupiter and Saturn. For the outer disk radii to set the scale of the present regular satellite systems, the solids from which the satellites formed must have been small enough to be coupled to the gas disk. This could occur if small solids are continually being supplied from the solar nebula or through collisional processes within the proto-satellite disk. However, if the solids grow on a timescale that is much shorter than the timescale for orbital decay due to gas drag, then the solids will decouple from the gas and the radial extent of the resulting satellite system would be set by the centrifugal radius, $r_{\rm c}$. The location and transport of solids in an actively supplied, photoevaporating circumplanetary disk is an important issue that merits further investigation. 

The constant outward mass flux in the outer regions of the circum-Jovian disk, presented in Figure \ref{fig:mass_transfer}, indicate that while very low surface densities exist in the outer regions of these disks there is still significant amounts of gas being transported to this region (see Figure \ref{fig:mass_transfer}). Assuming that solids are carried along with the gas as it is transported outward, there is sufficient mass present for satellite formation. Given a solar abundance of solids, we calculated that there is nearly twice as much mass in solids transported outward over a $10^5$ yr time period than is needed to form Callisto. There are reasons to believe that the solids-to-gas mass ratio would be higher than solar and therefore we take this value as an underestimate. We plan to investigate the actual distribution and transport of solids in the near future with a more comprehensive model that is currently in development.

Another aspect of our simulations that may have consequences on satellite formation is the density enhancement seen in our simulations at $\sim 25\ r_{\rm p}$, the location of peak mass infall. A similar density enhancement was seen in \citet{machida08}, but was not present in more recent simulations which include radiative transfer \citep{ayliffe09}. It is unclear at this time if these density enhancements would be present in a more realistic model in which the viscosity was determined locally. This is an interesting question and one which we plan to investigate in the near future. If it is real, these enhancements would have a significant impact on satellite formation. Density enhancements such as these are accompanied by pressure maxima. It has been shown that migrating solids can be trapped in such pressure maxima and rapidly grow into satellitesimals \citep{kretke09}.

In an effort to test what effect the location at which infalling material intersects the disk has on steady-state disk morphology, we have performed one test simulation in which the peak of the infalling material occurred at $35\ r_{\rm p}$ rather than at $25\ r_{\rm p}$. In the test simulation, the location of the disk outer edge was shifted farther out by $\sim 3 \%$ and the total disk mass increased by $\sim 5\%$. The changes are a result of a greater fraction of the infalling mass being transported outward rather than inward. This test indicates that it is important to identify the exact location at which the infalling mass accretes onto the circumplanetary disk. However, the current 3-D hydrodynamical simulations used to model infall from the solar nebula onto circumplanetary disks have insufficient resolution to identify the location exactly. We are satisfied that our treatment is sufficient for this study, yet we plan on including more precise results as they become available.

While the strength of the viscosity plays no role in the location of a disks outer boundary, it does play a significant role in the total mass contained in a given disk. In a steady state, the mass accretion rate, $\dot{M}\propto \nu \Sigma$, is constant. A constant mass accretion rate implies that the mass surface density must be proportional to the inverse of the viscosity. In our models, this means that the surface density is inversely proportional to the viscosity parameter, $\alpha$. One would naively assume that a larger surface density would result in larger satellites, but in actuality the opposite is true due to the increased rate of migration. A more massive, lower viscosity disk results in a less massive satellite system. \citet{canup02}find that satellites will only survive against type I migration for values of $\alpha \geq 10^{-3}$, assuming a solar gas-to-solid ratio for the infalling material. 

By varying the viscosity parameter, $\alpha$, we are able to produce a wide variety of total disk masses in our models. These masses range from $2.3\times 10^{-6}-1.9\times 10^{-4}\ M_{\rm S}$ in the Saturnian system to $5.5\times 10^{-6}-5.4\times 10^{-4}\ M_{\rm J}$ in the Jovian system. These radii are all many of orders of magnitude smaller than the $\sim 0.02\ M_{\rm J}$ inferred for the ``MMSN'' approach. Such low mass surface densities are a result of the $5\times 10^6$ yr timescale over which the infall occurs. Again, these conditions are necessary to produce the ice-rich compositions of the regular satellites of Jupiter and Saturn as well as the incomplete differentiation of Callisto and Titan.

The location of the disk outer boundary in our models depends only on the amount of FUV flux, as parameterized by our envelope temperatures. These radii are independent of the viscosity. This may not however be the case in a more realistic case where the viscosity is dependent on the local properties in the disk and is not simply proportional to the radius. We are currently developing a more sophisticated model in which the viscosity will be calculated locally. We seek to explore the morphology of steady-state disks with more realistic viscosities in the near future.

As discussed earlier, Jupiter, and to some extent, Saturn, is expected to open a gap in the solar nebula as a result of resonant interactions. Whether the gap opened be complete or partial, it would seriously inhibit the flow of gas onto circumplanetary,  satellite forming disks. The simulation presented in Figure \ref{fig:decay} shows that if such a restriction in infall is accompanied by a rapid dispersal mechanism, such as photoevaporation, it will cause the rapid removal of the circumplanetary disk as well. The rapid removal of gas would dynamically ``freeze'' the Galilean satellites in place. The rapid dispersal may also be aided by the increased ionization fraction, and increased viscosity, that may occur as a result of lower surface densities. We hope to investigate this in the future with a model that includes the local evaluation of viscosity. 

The partial, or incomplete, gap opened by Saturn would potentially have very different consequences \citep{sasaki10}. Again, photoevaporation would limit the size of the Saturnian subnebula based on the waning infall rate. The very low surface densities that would be present as a result of the waning inflow would be consistent with the incomplete differentiation of Titan and the formation of Saturn's rings \citep{barr10,canup10}. We plan a more thorough investigation of the decay of infall in the near future.

We feel that the gap-opening scenario described in \citet{sasaki10} can be further refined by including the effect of the inside-out clearing of the solar nebula that occurs in many simulations which include irradiation from a central source. In particular, we would like to draw the reader's attention to Figure 4 in \citet{gorti09}. The snap-shots of the mass surface density in Figure 4 show the photoevaporation front sweeping outward past Jupiter's location rather rapidly, but taking much longer to sweep past Saturn's location. Although it is difficult to determine from the snap-shots shown in Figure 4, the outward-traveling photoevaporation front is moving at approximately $8.8\ {\rm AU}\ {\rm Myr}^{-1}$ through the region where Jupiter is located. This is roughly twice as fast as it is moving through the region in which Saturn is located, which is approximately $4.5\ {\rm AU}\ {\rm Myr}^{-1}$. The outward-traveling front not only sweeps past Saturn at a slower rate than Jupiter, but nearly $1\ {\rm Myr}$ later as well.

The inside-out clearing of the solar nebula may have exposed Jupiter's circumplanetary disk to much more solar flux at an earlier time than Saturn. Also, at Saturn's location the surface density seems to be decreasing globally on a similar timescale over which the front moves outward. If so, it may mean that the shutoff of the infall from the solar nebula onto the Jovian subnebula, because of the local depletion of the solar nebula due to inside-out clearing, would have happened sooner and on a shorter timescale than in the Saturnian system. In our scenario, the slow infall rates required by \citet{canup02,ward10} would occur as a result of gap opening, or partial gap opening in the solar nebula. Slow infall from the solar nebula, onto the circumplanetary nebulae, would continue until the outward-traveling photoevaporation front passes their respective locations, globally clearing the solar nebula and ceasing any further infall. In this scenario, the final structure of the regular satellite systems of the giant planets that we see today would have been determined by the rate and timing of the inside-out clearing of the solar nebula. 

On a final note, the processes responsible for the transport and removal of angular momentum in astrophysical disks is highly uncertain. It is even more uncertain in the case of circumstellar and circumplanetary disks where the ionization fraction of the gas may be low enough to prevent magnetorotational instabilities from occurring. Whatever the mechanism for angular momentum transport, it must have occurred in circumplanetary disks to enable the accretion of the planet. Gas liberated through photoevaporation carries angular momentum away with it and may thereby provide a natural explanation for its removal from the system.

%%%%%%%%%%%%%%%%%%%%%%%%%%%%%%%%%%%%%%%%%%%%%%%%%%%%%%%%%%%%%%%%%%%%%%%%%%%%%%%%%%%%%%%%%%%%%%%%%%%%%%%%%%%%%%%

\acknowledgements

This work was supported by NASA Earth and Space Science Fellowship \#10-Planet10F-0057 and by the Cassini Project. G.R. Stewart was also partly supported by NASA's Planetary Geology and Geophysics Program.



%\begin{thebibliography}{}
\bibliography{bibliography.bib}
\bibliographystyle{apj}
%\end{thebibliography}

\clearpage

\end{document}